\documentclass[12pt]{article}
\usepackage[final]{graphicx}
\usepackage{amsmath}
\usepackage{amssymb}
\usepackage{xcolor}
\usepackage{tensor}
\usepackage{braket}
\usepackage{units}
\usepackage[hypertex,colorlinks=true,linkcolor=red,citecolor=blue]{hyperref}

\usepackage{amssymb}
\newcounter{comment}

{\refstepcounter{comment}%
\begin{quote}
\ttfamily\small$\blacksquare$ \textbf{\underline{Comment} $\sharp$\thecomment:}}%
{\end{quote}}

{
\begin{quote}
\ttfamily\small$\blacktriangleright$ \textbf{\underline{Reply} $\sharp$\thecomment:}}%
{\end{quote}}

\newcommand{\GeV}{{\rm GeV}}

\def\muFgpd{\relax\ifmmode\mu_\text{F,GPD}^2\else{$\mu_\text{F,GPD}^2${ }}\fi}
\def\muFda{\relax\ifmmode\mu_\text{F,DA}^2\else{$\mu_\text{F,DA}^2${ }}\fi}
\def\muO{\relax\ifmmode{\mu_{0}^{2}}\else{$\mu_{0}^{2}${ }}\fi}
\def\Mev{\relax\ifmmode{\text{MeV}}\else{MeV{ }}\fi}

\def\Li{\relax\ifmmode{\text{Li}_{2}}\else{Li$_2${ }}\fi}
\def\im{\Im{\rm m}}
\def\re{\Re{\rm e}}

\font\cmss=cmss12 
\def\1{\hbox{{1}\kern-.25em\hbox{l}}}
\def\bfZ{\relax{\hbox{\cmss Z\kern-.4em Z}}}
\begin{document}

\begin{titlepage}

\centerline{\large \bf Double distributions and generalized parton distributions from the}
\centerline{\large \bf parton number conserved light front wave function overlap representation}

\vspace{10mm}

\centerline{D.~M\"{u}ller}

\vspace{4mm}
\centerline{\it Theoretical Physics Division, Rudjer Bo{\v s}kovi{\'c} Institute}
\centerline{\it HR-10002 Zagreb, Croatia}

\vspace{10mm}

\begin{abstract}
\noindent
We show that Mellin moments of generalized parton distributions, given as even polynomials in the skewness parameter, are obtained from the Taylor expansion of light front wave functions.  Furthermore, we derive non-standard versions of the inverse Radon transform to obtain the double distribution from the  parton number conserved light front wave function overlap. These transformations are utilized to extend a generalized parton distribution from the outer region to the central one. We exemplify the formalism for a  light front wave function that arises  from an AdS/QCD duality conjecture.
\end{abstract}
\vspace{0.5cm}

\noindent

\vspace*{12mm}



\end{titlepage}

\newpage


\section{Introduction}

Generalized parton distributions (GPDs) \cite{Mueller:1998fv,Radyushkin:1996nd,Ji:1996nm} are an important tool to study the internal structure of the nucleon in hard exclusive processes, which allow to get a phenomenological handle on the nucleon spin sum rule \cite{Ji:1996ek} or the transverse distribution of partons \cite{Ralston:2001xs,Burkardt:2002hr}.  They are defined as off-forward matrix elements of light--ray operators or, alternatively, as the overlap of light front wave functions (LFWFs) \cite{Diehl:1998kh,Brodsky:2000xy,Diehl:2000xz} and are rather intricate functions, depending on the longitudinal momentum fraction $x$, the so--called  skewness parameter $\eta$, the momentum transfer in the $t$--channel, and the factorization scale.

A flexible GPD analyze tool has been developed on the basis of a conformal partial wave expansion \cite{Kumericki:2007sa,Mueller:2013caa}, which is equivalent to the dual parametrization \cite{Polyakov:2002wz,Muller:2014wxa}, and so far experimental data could be described with a rather simple GPD model \footnote{There is a $\sim2\sigma$ problem with respect to new deeply virtual Compton scattering data from the HALL A collaboration at lower electron beam energy \cite{Defurne:2017paw}, which can be probably overcome by updating the GPD model or taking into account kinematical twist and higher order corrections \cite{Braun:2011dg,Braun:2014sta,Braun:2017cih}.}.   Unfortunately, a GPD based analysis of experimental data is highly model biased, and to reduce the degrees of freedom, it would be desired to incorporate in GPD models the two basic properties, which arise from Poincar{\'e} covariance and from the positivity of the norm in the Hilbert space of states.

The former one yields the polynomiality condition for Mellin moments of GPDs while the later one yields positivity bounds for the GPD in the outer region $|\eta| < |x|$. The polynomiality condition is explicitly implemented in the so--called double distribution representation \cite{Mueller:1998fv,Radyushkin:1997ki}, in which the GPD
is the Radon transform of the double distribution (DD) \cite{Teryaev:2001qm}. The polynomials are given in terms of form factors and generalized form factors (GFFs), which can be measured in Lattice QCD simulations, see, e.g., \cite{Hagler:2009ni}.

On the other hand, positivity bounds are explicitly manifested in the parton number conserved LFWF  overlap representation.
Since in QCD the bound state problem remains unsolved, LFWFs are often modelled \cite{Boffi:2007yc}, determined in a Fock state truncated dynamical approach, or obtained from an AdS/QCD duality conjecture \cite{Brodsky:2014yha}.  To restore the full GPD also the parton number changing LFWF overlap is needed. Since of truncation, this overlap is often not calculated and so the complete GPD remains unknown, i.e.,  it  cannot be shown that the polynomiality condition holds.

Often model builders consider it as a dilemma that polynomiality conditions and positivity bounds cannot be simultaneously guaranteed. However, let us remind that a general
integral representation in which both properties are satisfied has been given for some time in Ref.\ \cite{Pobylitsa:2002vi} and that maps from the outer GPD region to the central ones are known \cite{Mueller:2005ed,Kumericki:2008di}, e.g., by an inverse Mellin transform. Also the DD--representation has been obtained in the study of the parton number conserved overlap of certain two--particle LFWFs \cite{Hwang:2007tb} and their phenomenological uses was discussed in \cite{Muller:2014tqa}. Here, it was believed that the interplay of longitudinal momentum fraction  and transverse momentum dependence play a crucial role to establish the polynomiality condition. Also numerical methods are utilized nowadays to map the  parton number conserved LFWF overlap to the DD \cite{Chouika:2017dhe}.

In the following we utilize the parton number conserved LFWF overlap, which provides the GPD in the outer region, to 
calculate GFFs, DDs, GPDs, and also generalized distribution amplitudes (GDAs).
Thereby, we will consider for simplicity the GPDs of a (pseudo)scalar hadron, however, the results can be easily adopted for nucleon GPDs.
In Sec.\ \ref{sec-GFFs} we introduce our conventions, argue that initial and final state momentum fraction dependence is separable, and define the GFFs directly in terms of the parton number conserved LFWF overlap representation, i.e., from the GPD in the outer region.  In Sec.\ \ref{sec-DDrep} the DD representation is derived from the overlap representation, where no constraints among longitudinal and transversal degrees of freedom are taken into account. Various versions of the inverse Radon transform are presented. In Sec.\ \ref{sec-duality} we provide the GPD duality map from the outer region to the central one and use crossing to give also the mapping to GDAs. In Sec.\ \ref{sec-example} the formalism is exemplified for a two--particle LFWF, obtained from an AdS/QCD conjecture. Finally, we summarize and conclude.

\section{Definitions and generalized form factors}
\label{sec-GFFs}

In the following we choose a symmetric reference frame in which the incoming and outgoing hadron have light front  momenta
$$P^{\rm i} = \left(\frac{1+\eta}{2}P, \frac{2M^2+{\bf \Delta}_{\perp}^2/2}{(1+\eta)P}, -\frac{1}{2}{\bf \Delta}_{\perp} \right) \mbox{  and  }
P^{\rm f} = \left(\frac{1-\eta}{2} P, \frac{2M^2+{\bf \Delta}_{\perp}^2/2}{(1-\eta)P}, \frac{1}{2}{\bf \Delta}_{\perp}\right ),$$
respectively, where $M$ is the mass of the hadron. The transverse momentum ${\bf \Delta}_{\perp}$ can be expressed by the momentum transfer $t=(P^{\rm f}-P^{\rm i})^2$ and the skewness parameter $\eta$
as ${\bf \Delta}_\perp =\sqrt{(1-\eta^2)(t_0-t)}\,{\bf e}_\perp(\varphi)$ with the kinematical boundary value $t_0 = -\frac{4 M^2 \eta^2}{1-\eta^2}$ and the unit vector ${\bf e}_\perp(\varphi)=(\cos\varphi,\sin\varphi)$.
In the following we denote the initial and final momentum fractions of the struck parton as $x^{\rm i}= \frac{x+\eta}{1+\eta}$ and $x^{\rm f}= \frac{x-\eta}{1-\eta}$, respectively, and restrict $\eta$ to  non-negative values.

A (twist-two) GPD $F(x,\eta,t)$, we drop here and in the following the factorization scale dependence, might be separated in the parton ($-\eta \le x\le 1$) and anti-parton ($-1 \le x\le \eta$) contributions in such a manner that the polynomiality conditions holds true for both of them. Since the anti-parton contribution can be mapped to the region $-\eta \le x\le 1$, it is sufficient to 
consider a generic parton GPD, which has the support $-\eta \le x\le 1$ and is built from  scalar, fermionic, or bosonic fields.
The  moments of such a GPD are even polynomials in $\eta$ of order $n+2s$,
  \begin{equation}
\label{F-moments}
F_n(\eta,t)= \int_{-\eta}^1\!dx\,x^n F(x,\eta,t) =\sum_{m=0 \atop {\rm even}}^{n+2s}F_{nm}(t) \eta^m\,,
\end{equation}
where $s\in \{0,1/2,1\}$ is the spin of the partons and the GFFs are denoted as $F_{nm}(t)$. 
The positivity bounds for the GPD are solved by the Fourier transform of a positive definite quadratic form in the impact parameter (${\bf b}_\perp$) space \cite{Pobylitsa:2002iu},
 \begin{equation}
\label{F-over}
F(x\ge \eta,\eta,t) =\int\!\!\!\!\int\!\!d^2{\bf b}_\perp e^{i \frac{1-x}{1-\eta^2}{\bf b}\cdot{\bf \Delta}_\perp } \frac{(1-x)^{2s}}{1-x}
 \sum_N \phi_N\!\left(x^{\rm i},{\bf b}_\perp\!\right)\phi_N^\ast\!\left(x^{\rm f},{\bf b}_\perp\!\right)\,.
\end{equation}
This solution can be derived from the parton number conserved LFWF overlap representation,
where  the sum runs over all spectators, includes their corresponding quantum numbers, and contains also the integrations over their momentum fractions and transverse momenta.

Let us argue that the LFWF overlap representation (\ref{F-over}), which contains  the Fourier exponent $ e^{i \frac{1-x}{1-\eta^2}{\bf b}\cdot{\bf \Delta}_\perp }$, might be written in a form where the  initial and final momentum fractions of the struck parton are separated. This step is not necessary for deriving our  results, however,  it makes the discussions more transparent.  Expressing the $\eta$--dependence in the transverse momentum by  $x^{\rm i}$ and $x^{\rm f}$, it  reads as follows
\begin{eqnarray}
\label{Delta-sep}
\frac{1-x}{1-\eta^2}{\bf \Delta}_\perp = M \sqrt{-(1- x^{\rm i}+1- x^{\rm f})^2+4(1- x^{\rm i})(1- x^{\rm f})\left[1-\frac{t}{4 M^2}\right]}\,  {\bf e}_\perp (\varphi).
\end{eqnarray}
The separation of $x^{\rm i}$ and $x^{\rm f}$ dependence in the Fourier exponent can be achieved, at least in principle. For instance, 
in the chiral limit, $M\to 0$, the expression  (\ref{Delta-sep}) takes a much simpler form
$
\frac{1-x}{1-\eta^2}{\bf \Delta}_\perp = \sqrt{-(1- x^{\rm i})(1- x^{\rm f})t}\,  {\bf e}_\perp (\varphi).
$
For exponentially decreasing LFWFs one can utilize a power expansion $ e^{i \frac{1-x}{1-\eta^2}{\bf b}\cdot{\bf \Delta}_\perp }= \sum_{m=0}^\infty \frac{1}{m!}  \sqrt{1- x^{\rm i}}^m \sqrt{1- x^{\rm f}}^m  (i \sqrt{-t} {\bf e}_\perp (\varphi)\cdot {\bf b}_\perp )^m$, which allows us to cast the LFWF  overlap  (\ref{F-over}) into the form of an auxiliary 
 overlap representation,
\begin{equation}
\label{F-overt}
F^{\rm out}(x,\eta,t) =F(x\ge \eta,\eta,t) =(1-x)^{2s}\frac{ \sum_{N} \phi_{N}(x^{\rm i}|t)\phi_N^\ast(x^{\rm f}|t)}{1-x}\,.
\end{equation}
Here, the symbolic sum over $N$ contains also the summation over $m$, the integration over ${\bf b}_{\perp}$, and includes the corresponding expansion coefficients. For power--like decreasing LFWFs one might first perform the Fourier expansion w.r.t.\  $\varphi$ and represent the sum over $m$ by an integral over complex valued $m$.  The $M> 0$ case might be treated in a similar manner, where the separation is reached by utilizing additional sums/integrals, however,  this yields to cumbersome representations.
Generally, since of the additional measure that is included in the summation over $N$, the auxiliary overlap (\ref{F-overt}) is  not a positive definite form and it should also hold true for GPDs that violate positivity bounds.

Let us now define the GFFs $F_{nm}(t)$, appearing in the $\eta$--expansion of the GPD moments  (\ref{F-moments}),  by means of  the auxiliary overlap (\ref{F-overt}).
Assuming that  $\phi_N(x^{\rm i}|t)$ are smooth (or generalized) functions for $0<x<1$ and exploiting the chain rule,  where
$d x^{\rm i}/d\eta  = (1-x)/(1+\eta)^2$ and $d x^{\rm f}/d\eta  = - (1-x)/(1-\eta)^2$, their Taylor expansion in the vicinity of   $\eta=0$ can be written in terms of derivatives w.r.t.\ $x$
\begin{eqnarray}
\label{phi-expansion}
\phi_N(x^{\rm i}|t)= \sum_{k=0}^\infty \eta^k  \sum_{l=0}^k c^{kl}_N  (1-x)^l\phi^{(l)}_N(x|t).
\end{eqnarray}
Here, $c^{kl}_N$ are combinatorial coefficients and the derivatives $\phi^{(l)}_N(x|t)= \frac{d^l}{dx^l} \phi_N(x|t)$ are decorated by a factor $(1-x)^l$. Therefore, we can also expand
 the GPD in the outer region  (\ref{F-overt}), where the Mellin transformed Taylor coefficients  are GFFs
\begin{eqnarray}
\label{F_{nm}(t)}
F^{(s)}_{nm}(t)
 =\int_0^1\! dx \frac{x^n}{m!}\frac{d^m}{d\eta^m}F^{\rm out}(x,\eta,t)\Big|_{\eta=0}
 =
 \int_0^1\! dx \frac{ x^n (1-x)^{2s}}{m!} \frac{d^m}{d\eta^m} \frac{\sum_{N} \phi_{N}(x^{\rm i}|t)\phi_N^\ast(x^{\rm f}|t)}{1-x}\Big|_{\eta=0}.
\end{eqnarray}
Note that an equivalent definition of GFFs in terms of a GPD, e.g., see (21) in \cite{Kumericki:2008di},  was derived by means of both the operator product expansion and a dispersion relation.
The $m${\it th} derivative w.r.t.\ $\eta$ can be expressed in terms of derivatives w.r.t.\ $x$, see (\ref{phi-expansion}), where the highest possible one is $(1-x)^{m-1} \frac{d^m}{dx^m} \left[\phi_N(x|t) \phi^{\ast}_N(x|t)\right]$.  Furthermore,  $(1-x)^{s-\frac{1}{2}}\phi_N(x|t)$ are considered in the interval $[0,1]$ as quadratic integrable functions, which ensures the convergence of the integral (\ref{F_{nm}(t)}) at the upper integration limit $x=1$.  The overlap behaves at small $x$ as $\sim x^{-\alpha}$ with the restriction $\alpha < 1$ (reggeon exchange) for even $n$ and $\alpha  \sim 1$ (pomeron exchange) for odd $n$.
To ensure the convergence of the integral (\ref{F_{nm}(t)}) at the lower integration limit $x=0$, the highest possible $m$ value is restricted to $n$ ($n-1$)  for even (odd) $n$.

The GFFs with $m>n$ might  be constructed by expanding the pre--factor $(1-x)^{2s}$, e.g., for quark GPDs ($s=1/2$) we define for odd $ n$
\begin{eqnarray}
\label{F_{n+1n}(t)}
F^{(1/2)}_{nn+1}(t|) = -F^{(0)}_{n+1,n+1}(t)
= - \int_0^1\! dx\, x^{n+1}  \frac{1}{(n+1)!} \frac{d^{n+1}}{d\eta^{n+1}} \frac{\sum_{N} \phi_{N}(x^{\rm i}|t)\phi_N^\ast(x^{\rm f}|t)}{1-x}\big|_{\eta=0}
.
\end{eqnarray}
This GFF can be associated with the so--called $D$--term contribution, which was introduced to complete the polynomiality condition \cite{Polyakov:1999gs}.  Also in the presence of pomeron exchange the integral (\ref{F_{n+1n}(t)}) does not converge and to get a finite answer one might utilize analytical (canonical) regularization \cite{GelShi64}, see example in Sec.\ 5.3.3 of \cite{Muller:2014tqa}.

Gluonic GFFs $F_{nm}$ are commonly defined w.r.t.\  $x^{n-1}$  moments with $n\ge 1$ rather than $x^n$ ones, which implies the restriction $m \le n+2s-1= n+1$, i.e., they are reduced to the $s=1/2$ case.
If we stick here for brevity to the definition (\ref{F-moments}),  we have  $F^{(1)}_{nn+1}(t) = -2F^{(0)}_{n+1,n+1}(t|0) $ for odd $n$ and the redundant relation  $F^{(1)}_{nn+2}(t)=F^{(0)}_{n+2,n+2}(t)$ for even $n$.

The difference of the GFFs  (\ref{F-moments}), obtained from the  moments of the GPD, and the GFFs (\ref{F_{nm}(t)}), calculated from the GPD in the outer region, reads
\begin{eqnarray}
\label{Delta F_{nm}(t)}
 \Delta F_{nm}(t) = \lim_{\eta\to 0} \frac{1}{m!}\frac{d^m}{d\eta^m}  \left[\int_{-\eta}^\eta\!dx\,  x^n   F(x,\eta,t) -\int_0^\eta \!dx\,  x^n F^{\rm out}(x,\eta,t)\right],
\end{eqnarray}
where we have extended the support of $ F^{\rm out}(x,\eta,t)$, e.g., by Taylor expansion in the vicinity of $x$,  and  utilized that the limit $\eta\to 0$ and the integration over $x$ can be interchanged,  $$ \lim_{\eta\to 0} \frac{d^m}{d\eta^m} \left[\int_{\eta}^1 + \int_{0}^\eta \right]\!dx\,  x^n   F^{\rm out}(x,\eta,t) =
\int_0^1 \!dx\,  x^n\frac{d^m}{d\eta^m} F^{\rm out}(x,\eta,t)\Big|_{\eta=0}.$$
Below we will show that $ \Delta F_{nm}(t) =0$ is self--consistent if the GPD is constructed from $F^{\rm out}$.

\section{DD representation from  LFWF overlap}
\label{sec-DDrep}

First, we derive the DD  representation from the parton number conserved  LFWF overlap.
 The method is based on the Laplace transform and generalizes that one utilized for certain two--particle LFWFs  \cite{Muller:2014tqa}. Inserting the Laplace transform
\begin{eqnarray}
\label{varphi}
\phi_N(x^{\rm i}|t) = \int_0^\infty\! d\lambda^{\rm i}\, \varphi_N(\lambda^{\rm i}|t) e^{-\frac{\lambda^{\rm i}}{1-x^{\rm i}}}
\end{eqnarray}
 into the  auxiliary  overlap (\ref{F-overt}), introducing the auxiliary variable $\lambda$ via $\int_0^\infty\!d\lambda\, \delta(\lambda-\lambda^{\rm i} -\lambda^{\rm f} ) =1$, and expressing   the double integral over $\lambda^{\rm i}$ and $\lambda^{\rm f}$ by the DD--variables $y$ and $z$,
\begin{equation}
\lambda^{\rm i} = \frac{1-x-z(1-\eta)}{2(1-y)} \lambda\,, \quad \lambda^{\rm f} = \frac{1-x+z(1+\eta)}{2(1-y)} \lambda\,,
\end{equation}
provides us the DD--representation for the GPD
 \begin{eqnarray}
\label{DD-repr}
F(x,\eta,t) = (1-x)^{2s} \int_0^1\! dy \int_{-1+y}^{1-y}\! dz\, \delta(x-y-z\eta) f(y,z,t).
\end{eqnarray}
After rescaling of $\lambda$ with $1-y$  the DD reads in terms of the Laplace transformed auxiliary functions   as follows
\begin{equation}
\label{h(y,z)}
f(y,z,t)=  \frac{1}{2}\int_0^\infty\! d\lambda\, \lambda\, e^{-\lambda} \sum_N \varphi_N(y^{\rm i} \lambda|t) \varphi^\ast_N(y^{\rm f} \lambda|t)
\end{equation}
where $y^{\rm i} = (1-y-z)/2$ and $y^{\rm f} = (1-y+z)/2$, i.e., $y=1-y^{\rm i}-y^{\rm f}$ and $z=y^{\rm f}-y^{\rm i}$.
Note that one can transform the DD representation (\ref{DD-repr}) to the standard one with $s=0$, which is nothing but a Radon transform \cite{Teryaev:2001qm}. For instance, for the $s=1/2$ case an additional $D$--term contribution appears, which  is projected out by a limiting procedure,
\begin{eqnarray}
\label{D-term}
D(x/\eta,t)= {\rm sign}(\eta) \theta(1-|x/\eta|)  d(x/\eta,t)\,, \quad d(x,t)=  \lim_{\eta\to\infty} F(x \eta,\eta,t)\,.
\end{eqnarray}
The odd  moments of this antisymmetric function are the GFFs $F_{nn+1}(t)$, see  (\ref{F_{n+1n}(t)}).  Note that contributions to the  $D$--term might also arise from contact terms, which are not contained in the parton number conserved LFWF overlap, see \cite{SemenovTianShansky:2008mp,Muller:2015vha}.

A direct map of the auxiliary overlap (\ref{F-overt}) to the DD follows if  the inverse Laplace transforms w.r.t.\  the variable $r^{}=1/(1-x)$, i.e.,
$$
\varphi_N(y^{\rm i}\lambda |t)= \frac{1}{2\pi i} \int_{c-i\infty}^{c+i\infty}\! dr^{\rm i}\,  e^{y^{\rm i}\lambda  r^{\rm i}} \phi_N\!\left(\!\!\frac{r^{\rm i}-1}{r^{\rm i}}\big|t\!\!\right) 
\mbox{ and }
\varphi^\ast_N(y^{\rm f}\lambda |t)= \frac{1}{2\pi i} \int_{c-i\infty}^{c+i\infty}\! dr^{\rm f}\,  e^{y^{\rm f}\lambda  r^{\rm f}} \phi^\ast_N\!\left(\!\!\frac{r^{\rm f}-1}{r^{\rm f}}\big|t\!\!\right) ,
$$
are inserted into (\ref{h(y,z)}) and  the integration over $\lambda$ is performed. Note that the complex conjugation operation does {\em not} act on the $(r^{\rm f}-1)/r^{\rm f}$  argument. 
The integration path is parallel to the imaginary axis and all singularities of $\phi_N\!\left(\!\frac{r-1}{r}\big|t\!\right)$ should lie on its left hand side. Since the auxiliary functions  might have brunch points at $x=0$ $(r=1)$ and $x=1$ $(r=\infty)$,  the real parts of the integration variables $r^{\rm i}$ and $r^{\rm f}$ are chosen to be one,
\begin{equation}
\label{h(y,z)-1}
f(y,z,t)= \frac{1}{2(2\pi i)^2}\int_{-i \infty}^{i \infty}\! d r^{\rm i}\!\!\int_{-i \infty}^{i \infty}\! d r^{\rm f}
\frac{
 \sum_N \phi_N(\frac{r^{\rm i}}{1+ r^{\rm i}}|t) \phi^\ast_N(\frac{r^{\rm f}}{1+r^{\rm f}}|t)
}{
\left(y^{\rm i} r^{\rm i} +y^{\rm f}r^{\rm f}-y \right)^2
}\,.
\end{equation}
 A realistic overlap should vanish at $r \to \infty$ with $|{\rm arg}(r)| < \pi/2$ (i.e., for $x\to 1$) so that we can close contours of the integration paths by semicircles that surround the first and forth quadrant  at  $r\to \infty$. Utilizing Cauchy theorem, it follows that only for positive  $y$ , $y^{\rm i}$, or  $y^{\rm f}$ the integral  does not vanish, i.e.,  the support of $f(y,z,t)$ is $|z| \le 1-y\le 1$, also shown in (\ref{DD-repr}).

 One integration in (\ref{h(y,z)-1})  can be performed by  Cauchy theorem, providing us the $y$--derivative of a single integral
 \begin{equation}
\label{h(y,z,t)-2}
f(y,z,t)= \frac{1}{2\pi i}\frac{d}{dy}\int_{-i \infty}^{i \infty}\! d r
\frac{
 \sum_N \phi_N(\frac{r}{1+r}|t)\left[ \phi^\ast_N(\frac{ y - y^{\rm i}r}{ y+ y^{\rm f}  - y^{\rm i}  r}|t)-\phi^\ast_N(\frac{2+r}{1+r}|t)\right]
}{
z \left(r + \frac{1+z}{z}\right)
}\,,
\end{equation}
where the pole at $r =- \frac{1+z}{z}$ will not contribute.
The singularities of the integrand are on both sides of the integration path: $\phi_N^\ast$  has a  possible cut on the real axis for $r\ge y /y^{\rm i}$, where its argument  becomes negative, reaches infinity at $r=(y+y^{\rm f}) /y^{\rm i}$, and is for $r>(y+y^{\rm f}) /y^{\rm i}$ larger than one. We might close  in  (\ref{h(y,z,t)-2})  the contour on the r.h.s.\ and  pick up so the singularities of  $\phi^\ast$. Assuming that the auxiliary functions have only a branch cut on the negative real $r$--axis,  we find after changing the integration variable $r$ by an SL(2,R) map to $\eta=\frac{r - (1 + r) y}{1 + z + r z}$ that the DD
\begin{equation}
\label{h(y,z,t)-3}
f(y,z,t)=-\frac{d}{dy}\int_{\frac{y}{1 - z}}^{\frac{1 - y}{z}}\! d\eta
\frac{\sum_N \phi_N(\frac{y+\eta  (1+z)}{1+\eta }|t)\left[ \phi^\ast_N(\frac{y-\eta  (1-z)}{1-\eta }+i\epsilon|t)-\phi^\ast_N(\frac{y-\eta  (1-z)}{1-\eta }-i\epsilon|t)\right]}{2\pi(1 - y - z \eta)}
\end{equation}
can be calculated from the imaginary part of the auxiliary functions, where the complex conjugation does not alter the $\epsilon$--prescription.  Note that for $\frac{-y}{1 + z} \le \eta \le  \frac{y}{1 - z} $ the argument of $\phi^\ast_N$ varies between $0$ and $1$ and, thus, the imaginary part vanishes. We might consider the auxiliary functions as holomorphic functions in the upper plane and its boundary value on the real axis are given by a Hilbert transform, denoted as $\textbf{H}$,
\begin{eqnarray}
\im \phi_N(x|t)=\textbf{H}(\phi_N)(x|t) =\lim_{\epsilon \to 0}\frac{1}{\pi}\int_{-\infty}^\infty dy\, \frac{(x-y)  \phi_N(y|t)}{(x-y)^2+\epsilon^2},
\end{eqnarray}
which requires to continue $\phi_N(x|t)$ on the whole real axis.

Clearly, by means of  (\ref{F-overt})  the DD can be also calculated from the GPD in the outer region, e.g., the analog of the integrals along the imaginary axis  (\ref{h(y,z,t)-2})  and real axis (\ref{h(y,z,t)-3}) read as follows
\begin{equation}
\label{h(y,z,t)-3F}
f(y,z,t)= \frac{-1}{2\pi i}\frac{d}{dy}\int_{-i \infty}^{i \infty}\! d r
\,
\left(\frac{1 - y + z}{1 + z + r z}\right)^{1-2 s}\,
\frac{F^{\rm out}\!\left(\frac{y + r z}{1 + z + r z},\frac{r - (1 + r) y}{1 + z + r z},t\!\right)}{z(r+\frac{1 + z }{ z})},
\end{equation}
and
\begin{equation}
\label{h(y,z,t)-4}
f(y,z,t)=-\frac{d}{dy}\int_{\frac{y}{1 - z}}^{\frac{1-y}{z}}\! d\eta
\,
\frac{\im F\!\left(y + z \eta ,\eta,t\!\right)}{\pi(1 - y - z \eta)^{2 s}},
\end{equation}
respectively, where the imaginary part $\im F(x,\eta) = \frac{1}{2}\left[F(x,\eta+i\epsilon)-F(x,\eta-i\epsilon)\right]$,
which is antisymmetric in $\eta$, appears in the central region  $|x|\le \eta$.  The $(x,\eta)$ arguments of the GPD run from the cross-over line $(y/(1-z),y/(1-z))$  to $(1,(1-y)/z)$.
The imaginary part is defined as boundary value of a holomorphic function w.r.t.\  the $\eta$-variable in the upper plane and can be calculated by means of a Hilbert transform,
\begin{eqnarray}
\label{imF}
\im F(x ,\eta,t)= \textbf{H}(F^{\rm out})(x,\eta,t) =\lim_{\epsilon\to 0}\frac{1}{\pi}\int_{-\infty}^\infty d\eta^\prime\, \frac{(\eta-\eta^\prime)  F^{\rm out}(x,\eta^\prime,t) }{(\eta-\eta^\prime)^2+\epsilon^2},
\end{eqnarray}
which might be reduced by means of symmetry under $\eta\to-\eta$ to a principal value integral on the real positive axis. To calculate this integral, the GPD
$F(\eta \le x,\eta)$ has to be analytically continued to the interval $\eta \in[x,\infty]$. 

Finally, inserting the Hilbert transform (\ref{imF}) into the integral  (\ref{h(y,z,t)-4}) and shifting/renaming variables $\eta \to (x-y)/z$  and $\eta^\prime \to \eta$ yield
\begin{equation}
\label{h(y,z,t)-5}
f(y,z,t)=\frac{1}{\pi^2}\frac{d}{dy}\int_{\frac{y}{1 - |z|}}^{1}\! dx\, {\rm pv}\!\int_{-\infty}^\infty\! d\eta\,
\frac{F^{\rm out}(x,\eta,t) }{(1 - x)^{2 s} (y+|z| \eta-x)},
\end{equation}
where we utilized the symmetry under $z\to -z$ reflection.
For $s=0$ this integral transformation looks like an inverse Radon transform, except that the support differs and $F^{\rm out}(x,\eta,t)$ is the analytic continuation of the GPD in the outer region.

\section{Duality map and crossing}
\label{sec-duality}

The continuation of a GPD from the outer to the central region can be done directly.  Inserting  (\ref{h(y,z,t)-3F})   into the DD integral $F(|x| \le \eta,\eta,t) = \frac{(1-x)^{2s}}{\eta} \int_{0}^{\frac{x+\eta}{1+\eta}}\! dy\, f(y,z=(x-y)/\eta,t)$ and utilizing the derivative w.r.t.\ $y$  yields an integral representation, which can be extended,
\begin{eqnarray}
\label{GPDcomplete}
F(x,\eta,t)=- \frac{(1-x)^{2s}}{2\pi i} \frac{x+\eta}{1-x} \int_{-i \infty}^{i \infty}\! \frac{d r}{r-\frac{x+\eta}{1-x}}
\left(\frac{x+\eta}{x+\eta +r x}\right)^{-2 s}
\frac{F^{\rm out}\left(\frac{r x}{x+\eta +r x},\frac{r \eta }{x+\eta+r x},t\right)}{x+\eta +r x}.
\end{eqnarray}
Indeed, this integral represents the full GPD.
Except for the $r$--pole, for  $|x|>\eta$ the integrand is holomorphic on the r.h.s.\ of the integration path. Thus, from the position of the $r$--pole  one realizes  that for  $x>\eta$ the GPD in the outer region follows while for $x<- \eta$ the GPD vanishes. For $|x| < \eta$  the integrand has singularities which are located on both sides of the integration path.

For positive (negative) $x$ the GPD (\ref{GPDcomplete}) might be evaluated from the singularities on the r.h.s.\ (l.h.s.), e.g.,  if only a  discontinuity on the positive (negative) real axis exists, we find a dispersion like relations which after a SL(2,R) transformation, $r=\frac{(\eta+x) \rho }{1-x \rho }$, takes the form
\begin{eqnarray}
\label{GPDcen-3}
F(x,\eta,t) &\!\!\!=\!\!\! &
\frac{{\rm sign}(x)}{\pi}\im  \int_{0}^{\frac{1}{x}}\! d\rho  \frac{(1-x)^{2s}}{(1 -x \rho )^{2 s}}
\,
\frac{F^{\rm out}(x \rho  ,\eta \rho,t)}{1-\rho -i \epsilon}\,.
\end{eqnarray}
Some care is needed to evaluate this integral in the central region. For positive $x$ the Cauchy kernel $\frac{1}{1-\rho-i\epsilon} = i \pi \delta(1-\rho)+{\rm pv} \frac{1}{1-\rho}$ possesses an imaginary and real part, which yields
\begin{eqnarray}
\label{GPDcen-1}
F(0 < x \le \eta,\eta,t) &\!\!\!=\!\!\! &\re F^{\rm out}(x,\eta,t)+
\frac{1}{\pi} {\rm pv}  \int_{0}^{\frac{1}{x}}\!\!  d\rho\, \frac{(1-x)^{2s}}{(1 -x \rho )^{2 s}}\,
\frac{\im F^{\rm out}(x \rho  ,\eta \rho,t)}{1-\rho}\,,
\end{eqnarray}
where the real part $\re F^{\rm out}(x,\eta,t)= \frac{1}{2} \left[F^{\rm out}(x,\eta+i \epsilon,t)+F^{\rm out}(x,\eta-i \epsilon,t)\right]$  is given by the principal value prescription.  For negative $x$, there is no pole contribution from the Cauchy kernel and changing the integration variable $\rho\to -\rho$ and exploiting the antisymmetry of  $\im F^{\rm out}(x ,-\eta,t)=-\im F^{\rm out}(x ,\eta,t)$, the dispersion integral reads
\begin{eqnarray}
\label{GPDcen-2}
F(-\eta \le x < 0,\eta,t)  &\!\!\!=\!\!\! &
\frac{1}{\pi} {\rm pv}  \int_{0}^{-\frac{1}{x}}\!\!  d\rho\, \frac{(1-x)^{2s}}{(1+x \rho )^{2 s}}\,
\frac{\im F^{\rm out}(-x \rho ,\eta \rho,t)}{1+\rho}\,.
\end{eqnarray}
 The function (\ref{GPDcen-1}) defines the continuation of (\ref{GPDcen-2}) to positive $x$ in such a manner that the boundary conditions $F(-\eta,\eta,t)=0$ and $F(\eta,\eta,t)=F^{\rm out}(\eta,\eta,t)$ are valid for  $\im F^{\rm out}(\eta,\eta,t)=0$, i.e., the imaginary part of the function $\phi_N(-x|t)$ vanishes in the limit $x\to +0$. 
Note that  the continuation to positive $x$ can be often performed analytically.  

Let us show that the GFFs, entering the  moments of the full GPD (\ref{F-moments}),  and  the GFFs (\ref{F_{nm}(t)}), calculated from the GPD in the outer region,  are the same, i.e., their difference  (\ref{Delta F_{nm}(t)}) vanishes. Since the  $m=n+2s$ cases are reduced to $m=n$ ones, see (\ref{F_{n+1n}(t)}), it is sufficient to consider the cases  $m \le n$ with $s=0$. Plugging (\ref{GPDcen-1}) and (\ref{GPDcen-2}) into the difference (\ref{Delta F_{nm}(t)}),
one immediately sees that the real part contributions of $F^{\rm out}(x,\eta,t)$ cancel each other. After mapping the contribution (\ref{GPDcen-2}) from negative $x$ to positive one and rescaling the integration variable $x$ by
$\eta$ the difference (\ref{Delta F_{nm}(t)}) reads
   \begin{eqnarray}
 \Delta F_{nm}(t)  \sim    \lim_{\eta\to 0} \frac{d^m}{d\eta^m} \eta^{n+1} \int_{0}^1\!dx\,  x^n
\frac{{\rm pv}}{\pi}   \int_{0}^{1}\! dx^\prime
\,
\frac{  x^\prime (1+\sigma)+x \eta(1-\sigma)}{x^2 \eta^2 -x^{\prime 2}}  \im F^{\rm out}\!\left(x^\prime ,  x^\prime/x,t\right),
\nonumber
  \end{eqnarray}
where the signature $\sigma= (-1)^{n+1}$.
The $x^\prime$--integral,  containing the Cauchy kernel,  is sensitive to the small $x^\prime$ behavior of  $ \im F^{\rm out}\!\left(x^\prime ,  x^\prime/x,t\right) \sim x^{\prime -\alpha}$ and it exists if $\alpha < 2$ and $\alpha < 1$ for $\sigma=+1$ and  $\sigma=-1$, respectively. It provides us a function $g(x\eta,\eta,t) \sim \eta^{-\alpha}$, which for $x\in[0,1]$  inherits the small $x^\prime$ behavior of  $ \im F^{\rm out}\!\left(x^\prime ,  x^\prime/x,t\right)$. Hence, the integral $\int_0^1\!dx\,x^n g(x\eta,\eta,t) \sim \eta^{-\alpha}$ exists for $n\ge 0$ and   the difference $\Delta F_{nm}(t)  \sim    \lim_{\eta\to 0} \frac{d^m}{d\eta^m} \eta^{n+1-\alpha}\sim    \lim_{\eta\to 0} \eta^{n-m+1-\alpha}$ vanishes for $m\le n$
with $\alpha <1$ for even $n$ ($\sigma=-1$) and $\alpha < 2$ for odd $n$ ($\sigma=+1$), where $m$ is always even.

Finally, let us comment on generalized distribution amplitudes (GDAs) $f(x,\eta,W^2)$, which are non-perturbative amplitudes for the production of a two hadron state \cite{Mueller:1998fv,DieGouPirTer98}.  They  are the crossed version of GPDs, where $f(x,\eta,W^2)= (1/\eta)F(x/\eta,1/\eta,t=W^2)$, and  can be directly obtained from the DD (\ref{DD-repr}) or the integral representations (\ref{GPDcomplete}) and (\ref{GPDcen-3}).
For instance, the crossing of (\ref{GPDcen-3})  provides us after rescaling of the integration variable $\rho \to \rho \eta$
\begin{eqnarray}
\label{GDA-1}
\texttt{f}(x,\eta,W^2)  &\!\!\!=\!\!\! &\frac{{\rm sign}(x)}{\pi}\im  \int_{0}^{1/x}\! d\rho \frac{(1-x/\eta)^{2s}}{(1 -x \rho)^{2 s}}
\,
\frac{F^{\rm out}\!\left(x \rho ,\rho,W^2\!\right)}{1-\eta \rho- i\epsilon}\,.
\end{eqnarray}
For $x>\eta$ the pole at $\rho=1/\eta$ is located outside of the integration region and, thus, we find
\begin{eqnarray}
\label{GDA1}
\texttt{f}(x> \eta,\eta,W^2)  &\!\!\!=\!\!\! &\frac{1}{\pi} \int_{0}^{1/x}\! d\rho \frac{(1-x/\eta)^{2s}}{(1 -x \rho)^{2 s}}
\,
\frac{\im  F^{\rm out}\!\left(x \rho ,\rho,W^2\!\right)}{1-\eta \rho}\,,
\end{eqnarray}
where the integrand for  $x=1$  vanishes if $\im  F^{\rm out}(\rho ,\rho,W^2)=0$.
The continuation to $x<\eta$ might be done by utilizing the transformation $x\to -x$ and $\eta\to -\eta$
\begin{eqnarray}
\label{GDA-2}
\texttt{f}(x< \eta,\eta,W^2)  &\!\!\!=\!\!\! &\frac{1}{\pi} \int_{0}^{-1/x}\! d\rho \frac{(1-x/\eta)^{2s}}{(1 +x \rho)^{2 s}}
\,
\frac{\im  F^{\rm out}\!\left(-x \rho ,\rho,W^2\!\right)}{1+\eta \rho}
\end{eqnarray}
for negative $x$ values, which corresponds to the crossed version of  (\ref{GPDcen-2}). The continuation to positive $x < \eta$ follows from (\ref{GPDcen-1}),
\begin{eqnarray}
\label{GDA-3}
\texttt{f}(x< \eta,\eta,W^2)  &\!\!\!=\!\!\! & \re F^{\rm out}\!\!\left(\frac{x}{\eta},\frac{1}{\eta},W^2\!\right) +\frac{1}{\pi} \int_{0}^{1/x}\! d\rho \frac{(1-x/\eta)^{2s}}{(1 -x \rho)^{2 s}}
\,
\frac{\im  F^{\rm out}\!\left(x \rho ,\rho,W^2\!\right)}{1-\eta \rho},
\end{eqnarray}
or might be reached by analytic continuation of (\ref{GDA-2}). For $x=\eta$ the momentum fractions $x^{\rm i} = (1+x)/(1+\eta)$ and $x^{\rm f} = (1-x)/(1-\eta)$ are equated to one.  Since we assume that the LFWFs vanish at $x=1$,  it follows from the crossed LFWF overlap representation (\ref{F-over}) that $F^{\rm out}(x=1,\frac{1}{\eta},W^2)$  vanishes. Consequently,  (\ref{GDA1}) and (\ref{GDA-3}) approaching the same value at $x=\eta$, i.e., the GDA is continues in this point.

\section{AdS/QCD example}
\label{sec-example}

To illuminate the formalism, we consider the two--particle LFWF that is obtained from an AdS/QCD duality conjecture \cite{Brodsky:2014yha}
\begin{equation}
\label{phi^{AdS/QCD}}
\phi^{\rm AdS/QCD}(x,{\bf b}_\perp)= N^{-1/2} \sqrt{x(1-x)} e^{-\frac{\kappa^2}{2}{\bf b}_\perp^2 x(1-x)},
\end{equation}
where $N$ is a normalization factor. The $t$-dependent quark PDF ($\eta=0$) can be easily obtained and reads
as $\sim e^{\frac{t (1-x)}{4 x \kappa ^2}}$, i.e., it is a constant for $t=0$ or  $x=1$ and it is exponentially suppressed for $t<0$, in particular, in the small $x$--region.
The  $x=1$ behavior contradicts our assumption that the LFWF overlap vanishes at this end--point. Nevertheless, the full GPD can be analytically constructed for  the ${\bf b}_\perp=0$ and $t=t_0$ cases. Utilizing a subtraction procedure, it is possible to restore the $t$-dependent GPD numerically. Note that as in the case of other two--particle LFWFs the direct calculation of the particle number changing LFWF overlap requires additional information, namely, the corresponding one-particle  and three--particle LFWF.

\subsection{The ${\bf b}_\perp=0$ case}
First we consider the ${\bf b}_\perp=0$ case, which enters the GPD that is integrated over $\frac{1-x}{1-\eta^2}{\bf \Delta}_\perp$, see LFWF overlap representation  (\ref{F-over}).
The Laplace transform of the LFWF (\ref{phi^{AdS/QCD}}) reads
$$
\varphi^{\rm AdS/QCD}(\lambda,{\bf b}_\perp=0) =N^{-1/2}\left[ \frac{e^{\lambda }}{\sqrt{\pi } \sqrt{\lambda }}- {\rm erfi}(\sqrt{\lambda })\right],
$$
where ${\rm erfi}(x)= \frac{2}{\sqrt{\pi}} \int_0^x\! dt\, e^{t^2}$ is the imaginary error function.  Utilizing this representation, we calculate from (\ref{h(y,z)}) the DD in a straightforward manner,
\begin{eqnarray}
\label{h^{AdS/QCD}}
h^{\rm AdS/QCD}(y,z)=\frac{N^{-1}}{\pi \sqrt{(1-y)^2-z^2} } \left[
\frac{1}{2}-\frac{(y-z)\sqrt{1-y-z}}{\sqrt{1+y-z}}  {\rm arctanh}\,\frac{\sqrt{1-y-z}}{\sqrt{1+y-z}}+\{z\to -z\}\right].
\nonumber\\
\end{eqnarray}

Alternatively, one might start from the GPD in the outer region
\begin{eqnarray}
\label{H^{AdS/QCD}-out}
H^{\rm AdS/QCD}(x\ge \eta,\eta)=N^{-1}  (1-x)^{2s}\frac{\sqrt{x^2-\eta^2}}{1-\eta^2}
\end{eqnarray}
and its imaginary part, which can be also evaluated from the Hilbert transform (\ref{imF}),
$$
\im H^{\rm AdS/QCD}(x,\eta+i \epsilon)=-N^{-1}\theta(|\eta|-|x|) {\rm sign}(\eta)  (1-x)^{2s}\frac{\sqrt{\eta^2-x^2}}{1-\eta^2},$$
to calculate from (\ref{h(y,z,t)-4}) the DD (\ref{h^{AdS/QCD}}). Note that the poles at $\eta=\pm 1$ in the imaginary part are treated by the principal value prescription.
The Radon transform (\ref{DD-repr}) reproduces the  GPD  in the outer region, given in (\ref{H^{AdS/QCD}-out}), and yields its continuation
\begin{eqnarray}
\label{H^{AdS/QCD}-cen}
H^{\rm AdS/QCD}(|x| \le \eta,\eta)= \frac{(1-x)^{2s}}{N(1 - \eta^2)}
\frac{\sqrt{\eta^2 - x^2}}{2 \pi }\left[\ln\frac{(1-x)^2}{\eta^2  - x^2}-\frac{1}{\eta}\ln\frac{\eta  - x}{\eta+x}\right]
\end{eqnarray}
 into the central region. This result  can be easily derived from the dispersion integral (\ref{GPDcen-2}) for negative $x$ and its analytical continuation to positive $x$.

The  moments (\ref{F-moments}) of the full GPD, specified in (\ref{H^{AdS/QCD}-out}) and (\ref{H^{AdS/QCD}-cen}), are built by the  GFFs that are obtained from  the GPD in the outer region, see (\ref{F_{nm}(t)}) and (\ref{F_{n+1n}(t)}). They read as follows
$$
H^{\rm AdS/QCD}_{nm}=\frac{1+(-1)^m}{2 N}\frac{ \Gamma\!\left(\frac{m-1}{2}\right) \Gamma(2-m+n) \Gamma(2 s+1)
 }{
\Gamma\!\left(-\frac{1}{2}\right) \Gamma\!\left(1+\frac{m}{2}\right) \Gamma(3-m+n+2 s)}\,
{_4F_3}\left({1,-\frac{m}{2},\frac{2+n-m}{2},\frac{3+n-m}{2} \atop \frac{3-m}{2},\frac{3+2s+n-m}{2},\frac{4+2s+n-m}{2}}\Big|1\right)
$$
for $m\le n$ and
$$
H^{\rm AdS/QCD}_{n,n+1}=\frac{1-(-1)^n}{2 N} \frac{\Gamma\!\left(\frac{2+n}{2}\right)}{4\Gamma\!\left(\frac{1}{2}\right) \Gamma\!\left(\frac{5+n}{2}\right)}\left[S_1\!\left(\frac{n}{2}\right)-S_1\!\left(\frac{1}{2}\right)\right]
$$
for  $n=n+1$ and $s=1/2$ (quark case). Here, ${_4F_3}$ denotes a hypergeometric function and  $S_1(k)=\sum _{n=1}^{\infty } \frac{k}{n (n+k)}$ is a harmonic sum.

\subsection{GPD at $t=t_0$}

Next let us consider the AdS/QCD  GPD at $t=t_0$. It reads in the outer region,  see below (\ref{H^{AdS/QCD}-out_t}),  as follows
\begin{eqnarray}
\label{H^{AdS/QCD}-out_t0}
H^{\rm AdS/QCD}(x\ge \eta,\eta,t_0) = \frac{(1-x)^{2s}}{N (1-x)}\,\frac{ \sqrt{x^2-\eta^2}(1-\eta^2)}{x(1+\eta^2)-2\eta^2}\,.
\end{eqnarray}
In comparison to the $\frac{1-x}{1-\eta^2}{\bf \Delta}_\perp$ integrated version  (\ref{H^{AdS/QCD}-out}), the GPD at $t=t_0$ is enhanced in the $x\to 1$ limit by a factor $1/(1-x)$. This implies logarithmical divergent integrals either at $r=\pm i \infty$ or $x^\prime =1$, which are treated by subtraction, where the subtraction term is afterwards calculated with a cut-off $\sqrt{\Lambda^2-1}$ or
$1-1/\Lambda$.  The GPD  possesses also a pole at $x=2 \eta^2/(1+\eta^2) \le \eta$ in the central region. It will be treated in dispersion like integrals by the principal value prescription while no specific treatment is needed in integrals along the imaginary axis.  Since of these two difficulties, the DD  becomes a cumbersome function and it is more appropriate  to find the full GPD from  a duality map.

 The extension of the GPD (\ref{H^{AdS/QCD}-out_t0}) is straightforwardly achieved  for negative $x$ by means of (\ref{GPDcen-2})  and the analytical continuation to positive $x$. We decompose, e.g., the quark GPD, as
\begin{eqnarray}
\label{H^{AdS/QCD}-cen_t0}
H^{\rm AdS/QCD}(|x| \le \eta,\eta,t_0) &\!\!\!= \!\!\!& \frac{1}{2\pi N}\Bigg[
\sqrt{\eta^2 - x^2}\left\{
\frac{(1-\eta^2)}{x(1+\eta^2)-2\eta^2}\ln\frac{(1-x)^2}{\eta^2 - x^2} +\frac{1}{x}\,\ln\frac{x^2}{\eta^2 - x^2}
\right\}
\nonumber\\
&&\phantom{ \frac{1}{2\pi N}}+ \left\{
\frac{\eta(1 - 2 x + \eta^2)}{x(1+\eta^2)-2\eta^2}-\frac{\eta}{x}
\right\} \ln\frac{\eta-\sqrt{\eta^2 - x^2}}{\eta+\sqrt{\eta^2 - x^2}}
\Bigg]+ d\!\left(\!\frac{x}{\eta},t_0\!\!\right),\quad\quad
\end{eqnarray}
where we separated the $D$--term contribution,
\begin{eqnarray}
\label{d(x,t_0)}
d(x,t_0)&\!\!\!= \!\!\!&  \frac{1}{\pi N}\left[ - \frac{ \sqrt{1 - x^2}}{x} \left\{ \ln\frac{1}{\Lambda} + \ln\frac{x^2}{1-x^2}\right\}
  +\frac{1}{x}  \ln\frac{1-\sqrt{1- x^2}}{1+\sqrt{1 - x^2}}\right],
\end{eqnarray}
which divergences if the cut-off is removed, i.e., in the limit $1/\Lambda \to 0$. The extension procedure removed  the $x=2\eta^2/(1+\eta^2)$ pole in the central region, i.e., the two contributions on the r.h.s.\ of  (\ref{H^{AdS/QCD}-cen_t0}), which contain this pole, will cancel each other. However, note that
the GPD has a integrable $1/x$--pole with  residue $-\eta  \ln(2/\Lambda)/\pi$ and that the $D$--term does not absorb this pole rather it has there  the residue $-\eta\ln(4/\Lambda )/\pi$.

The moments of the resulting GPD, specified in (\ref{H^{AdS/QCD}-out_t0}),  (\ref{H^{AdS/QCD}-cen_t0}), and (\ref{d(x,t_0)}), are built by the GFFs that are obtained from the expansion (\ref{F_{nm}(t)}) of  $H^{\rm AdS/QCD}(x \ge \eta,\eta,t_0)$,
\begin{eqnarray}
H^{\rm AdS/QCD}_{n m}(t_0)
&\!\!\!=\!\!\!&  \frac{1+(-1)^m}{2N}\frac{\Gamma(\frac{m-1}{2})}{\Gamma(-\frac{1}{2})\Gamma(\frac{m+2}{2})} \Bigg[
\frac{\Gamma(1+n-m)\Gamma(2 s)}{\Gamma(1+2 s+n-m)}
-m \int_{0}^1\!dx\, x^{n}(1-x)^{2s}
\\
&&\phantom{ \frac{1+(-1)^m}{2N}\frac{\Gamma(\frac{m-1}{2})}{\Gamma(-\frac{1}{2})\Gamma(\frac{m+2}{2})} \Bigg[}
 \times   \frac{2 (2 - x)^ {\frac{m}{2}}}{(2-x)x^{m/2} }\; {_2F_1}\!\left({ \frac{2 - m}{2},- \frac{1}{2} \atop \frac{1}{2}}\Big| \frac{-(1 - x)^2}{x(2- x)}\right)
 \Bigg],
\nonumber
\end{eqnarray}
for $m\le n$ and from (\ref{F_{n+1n}(t)}) for the quark  ($s=1/2$) case,
$$
H^{\rm AdS/QCD}_{n,n+1}(t_0)=\frac{1-(-1)^n}{2 N} \frac{\Gamma\left(\frac{n}{2}\right)}{\Gamma\left(-\frac{1}{2}\right) \Gamma\left(\frac{3+n}{2}\right)}\left[ \ln\frac{1}{\Lambda} + S_1\!\left(\frac{n}{2}\right)-S_1\!\left(\frac{-1}{2}\right)\right].
$$

In the left panel of Fig.\ \ref{Fig-1} we show the resulting GPD for $\Lambda=2$, i.e., the $1/x$--pole is removed and a logarithmical divergency remains, $N=1$, and three values for $\eta$:  0 (solid), 0.5 (dashed), and 1 (dash--dotted).  Essentially, we constructed out of a horizontal line ($\eta=0$) a GPD which vanishes at the end point $x=-\eta$ and for $\eta<1$ also at the cross-over point $x=\eta$.

\subsection{Numerical treatment for $t$--dependent GPD}
The  $t$-dependent AdS/QCD GPD  in the outer region follows from the Fourier transform  (\ref{F-over}) within  the AdS/QCD  LFWF (\ref{phi^{AdS/QCD}}),
\begin{eqnarray}
\label{H^{AdS/QCD}-out_t}
H^{\rm AdS/QCD}(x\ge \eta,\eta,t) = \frac{1}{N}\frac{ \sqrt{x^2-\eta^2}(1-\eta^2)}{x(1+\eta^2)-2\eta^2} \exp\left\{{\frac{t(1-\eta^2)+4 M^2\eta^2}{4\kappa^2} \frac{1-x}{x(1+\eta^2)-2\eta^2} }\right\},
\end{eqnarray}
where the normalization $N$ absorbs now a factor $\kappa^2/2 \pi$.
Its continuation to the central region can be numerically achieved, however, there are obstacles. First, we observe that the extension of the GPD into the central region suffers now from an essential singularity at $x=2 \eta^2/(1+\eta^2) \le \eta$.
Thus, we work in the following with integrals in which the integration path is chosen along the imaginary axis. In addition the GPD  contains  for $x> \eta$ and $-t < -t_0$ an exponential enhancement factor, which corresponds to a violation of the stability condition for a two--particle bound state. In our case this violation implies instable  numerics and inconsistent results. This can be cured if the LFWF (\ref{phi^{AdS/QCD}}) is decorated with some exponential mass dependence, e.g.,
\begin{equation}
\label{phi^{AdS/QCD}-mod}
\phi^{\rm AdS/QCD}(x,{\bf b}_\perp) \to \phi^{\rm AdS/QCD}(x,{\bf b}_\perp) e^{-\frac{1-x}{x} \frac{M^2}{2\kappa^2}}\,,
\end{equation}
which guarantees that the argument of the resulting exponent,
\begin{eqnarray}
\label{H^{AdS/QCD}-out_t-M}
H^{\rm AdS/QCD}(x\ge\eta,\eta,t)  &\!\!\!=\!\!\!& \frac{1}{N}\frac{ \sqrt{x^2-\eta^2}(1-\eta^2)}{x(1+\eta^2)-2\eta^2}
\\&&\times
\exp\left\{\left(\frac{t-4 M^2}{4 \kappa ^2}\left(1-\eta ^2\right)-\frac{(1-x)^2 }{x^2-\eta ^2} \frac{ \eta ^2M^2}{\kappa ^2}\right)  \frac{1-x}{x(1+\eta^2)-2\eta^2}\right\},
\nonumber
\end{eqnarray}
remains negative for $t\le 4 M^2$. Note that the $x$--shape of the PDF/GPD drastically depends on the modification which has been chosen.

To get rid of the logarithmical singularity at $r=\pm i \infty$ we perform a two--step procedure. First we subtract the $D$--term contribution, which
is projected out by taking the limit as shown in (\ref{D-term}). The extension procedure for the quark GPD  in terms of the integral  (\ref{GPDcomplete}) reads then
\begin{eqnarray}
\label{H-D}
H(x,\eta,t)- D(x,\eta,t)=  \frac{-1}{2\pi i}\int_{-i \infty}^{i \infty}\! dr
\left[
\frac{H^{\rm out}(\frac{x r}{x + \eta+ x r}, \frac{\eta r}{x + \eta+ x r},t)}{
r-\frac{x+\eta}{1-x}
}-\frac{H^{\rm out}(\frac{x r}{x + \eta+ x r}, \frac{\eta r}{x + \eta+ x r},t)}{
r+\frac{x+\eta}{x}}
\right].
\end{eqnarray}
This integral provides a robust numerical treatment for the  GPD $H^{\rm AdS/QCD}$. The $D$--term is given by the antisymmetric function  $d(x,t)=-d(-x,t)$, and it is calculated from the integral
\begin{eqnarray}
\label{d(x,t)}
d(x,t)=  \frac{-{\rm sign}(x)}{2\pi i}\int_{-i \infty}^{i \infty}\! \frac{dr}{1+r}
\left[
H^{\rm AdS/QCD}\!\left(\frac{r}{1+ r},\frac{1}{|x|} \frac{r}{1+ r},t\!\right)-\frac{{\rm sign}(r/i)}{N}\sqrt{\frac{1-x^2}{x^2}}
\right]+ d^\Lambda(x),
\nonumber\\
\end{eqnarray}
where we have rescaled the integration variable $r\to r (1+x)/x$. The second term in the square brackets is the subtraction term $\lim_{r\to \pm i \infty}H^{\rm AdS/QCD}\!\left(\frac{r}{1+ r},\frac{1}{|x|} \frac{r}{1+ r},t\!\right)$, and adding it again gives the cut--off dependent part of the DD
$d^\Lambda(x)=\frac{\sqrt{1 - x^2}}{x}  \frac{\ln\Lambda}{\pi N}$, which coincide with that in (\ref{d(x,t_0)}). The GPD $H^{\rm AdS/QCD}(x,\eta,t)$  is numerically evaluated form (\ref{H-D}) and  (\ref{d(x,t)}). It contains a $ 1/x$--pole, however, its residue  depends on $M$ and $t$.

\begin{figure}[t]
\begin{center}
\includegraphics[width=17cm]{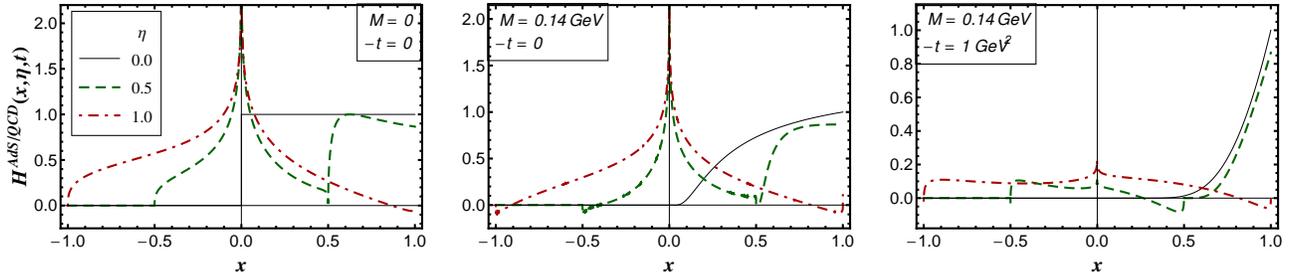}
\end{center}
\vspace{-5mm}
\caption{\small   The GPD $H^{\rm AdS/QCD}(x,\eta,t)$, evaluated from (\ref{H-D}) and (\ref{d(x,t)}),  for  different $\eta$--values: 0 (solid), 0.5 (dashed), and 1 (dash--dotted) versus $x$, where  $N=1$ and $\kappa=0.25\, \GeV$.   Left:
$t=t_0=0$ and  $\Lambda=2$.
Middle: $M=0.14\, \GeV$, $t=0$, and $\Lambda=2.326$.
Right:  $M=0.14\, \GeV$, $t=-1\, \GeV^2$, and $\Lambda=7.991$.
\label{Fig-1}
}
\end{figure}
For the  special case $t=t_0=0$, displayed in the left panel of Fig.\ \ref{Fig-1},  the analytic expression  (\ref{H^{AdS/QCD}-cen_t0}) agrees with numerical results.
In the middle and right panel of Fig.\ \ref{Fig-1} we show the GPD $H^{\rm AdS/QCD}$ for the modified LFWF ansatz (\ref{phi^{AdS/QCD}-mod}) with  $M =0.14\, \GeV$ and $\kappa =0.25\, \GeV$  at $t=0$ and $t=-1\, \GeV^2$, respectively, versus $x$ and three values of $\eta$:  0 (solid), 0.5 (dashed), and 1 (dash--dotted).  The  $ 1/x$--pole has been (nearly) removed by setting $\Lambda=2.326$ and  $\Lambda=7.991$ for $t=0$ and  $t=-1\, \GeV^2$, respectively. Comparing the left ($M=0$) and middle  ($M=0.14\, \GeV$)  panel one realizes that the shapes of the GPD are rather similar, except for the $\eta=0$ case, where  $H^{\rm AdS/QCD}(x,\eta=0,t=0)=e^{-M^2(1-x)/\kappa^2 x }/N$, and in the vicinity of the  cross--over and end points.
For increasing value of $-t$ the exponential suppression in the outer region increases,  the remaining singularity at $x=0$ diminishes, and the behavior at the cross--over and end points becomes steeper, see right panel. Finally, we have also numerically checked that the first few GPD moments (\ref{F-over}) are built by the GFFs (\ref{F_{nm}(t)}), which are calculated from the GPD in the outer region.

\section{Summary}

Starting from the parton number conserved LFWF overlap in off--forward kinematics, which corresponds to a GPD in the outer region,  we showed that one can calculate  GFFs directly from the  LFWFs.  It was also shown that these GFFs are those which are obtained from the moments of the full GPD. These results confirm previous ones, which were obtained in the operator product expansion framework. On the other hand, from the Laplace transform of axillary $t$--dependent  ``wave functions''  one can derive the DD representation for GPDs, which after crossing provides also GDAs. Thereby, the DD is given in terms of a double integral  along the imaginary axes, where, it is assumed that the singularities of the auxiliary functions  lie on the l.h.s.\ of the integration path.  From this transformation we derived single integral representations for the DD and the GPD in the central region, where the integration path goes along the imaginary axis or, alternatively, on the real axis.   While the former integral can be straightforwardly utilized for numerics, the latter one is a dispersion like relation which contains besides the pole of the Cauchy kernel also singularities that stem from the imaginary part of the  auxiliary overlap itself. Here, we also assumed that the GPD in the outer region can be considered as a holomorphic function in the upper $\eta$--plane. Under this assumption we derived a double integral transformation which is rather similar to the known formula for the inverse Radon transform.   In any case the analytic continuation of the parton number conserved overlap or  the GPD in the outer region is needed. The only ambiguity  arises from $D$--term contributions that are entirely related to the central GPD region, e.g., arising from contact terms.

In such a mapping approach the parton number changing LFWF overlap is (almost) uniquely restored from the parton number conserved one. However, this does not necessarily mean that models for LFWFs have proper behavior under Poincar{\' e}  transformations, e.g.,  they should satisfy the  bound state equation $P^- = (M^2+ {\bf P}_\perp)/P^+$. So far it is possible, covariance under Poincar{\' e}  transformations can be checked by verifying the polynomiality condition due to explicit LFWF calculations.   LFWF models might also violate the stability condition for the bound state, which can imply numerical  instabilities in the mapping approach and inconsistent results, i.e., the resulting GPD moments are not built by the GFFs that are calculated from the LFWFs.

Nevertheless, the mapping approach allows to build models for GPDs and GDAs  that satisfy both polynomiality conditions and positivity bounds and it provides a LFWF interpretation for the struck parton.  As advocated in \cite{Muller:2014tqa}, see also \cite{Chouika:2017dhe}, this can be used for setting up a unifying LFWF phenomenology for both hard exclusive and (semi)inclusive hadronic reactions. Thereby, it is natural  also to include ${\bf k}_\perp$--unintegrated parton distribution functions or even ${\bf k}_\perp$--unintegrated double distributions \cite{Muller:2014tqa}, where later are making contact to so-called Wigner functions.   However, building up such  phenomenological framework is challenging. For instance,  in a global GPD fitting procedure to experimental data a flexible parametrization of the GPD normalization, in particular on the cross-over line $x=\eta$, is needed, which sofar is only solved within the conformal partial wave expansion or, alternatively, the dual parameterization of GPDs. 

\section*{Acknowledgement}
For recent discussions I like to thank  G.~Duplan\v{c}i\'{c}, K.~Kumeri{\v c}ki, and  H.~Moutarde. 
This work has been supported by the NEWFELPRO grant agreement no.\ 54. 


\end{document}